\documentclass[twocolumn]{aastex62}
\usepackage{latexsym}
 \shorttitle{Superadiabaticity in giant planets} 
 \shortauthors{Debras, Chabrier \& Stevenson}

 \received{16/10/2020}
\accepted{03/05/2021}
\published{ - }
\usepackage{amsmath}

\def\apjs{Astrophys. J., Suppl. Ser. }
\def\apjl{Astrophys. J., Lett. }

\def\aap{Astronomy and Astrophysics}

\begin{document}
\title{Superadiabaticity in Jupiter and giant planet interiors}

\author{Florian Debras}
\affiliation{IRAP, Universit\'e de Toulouse, CNRS, UPS, Toulouse, France} 
\affiliation{Ecole normale sup\'erieure de Lyon, CRAL, UMR CNRS 5574, 69364 Lyon Cedex 07,  France} 
\affiliation{School of Physics, University of Exeter, Exeter, EX4 4QL, UK}

\author{Gilles Chabrier}
\affiliation{Ecole normale sup\'erieure de Lyon, CRAL, UMR CNRS 5574, 69364 Lyon Cedex 07,  France} 
\affiliation{School of Physics, University of Exeter, Exeter, EX4 4QL, UK}

\author{David J. Stevenson}
\affiliation{Division of Geological and Planetary Sciences, California Institute of Technology
1200 E California Blvd, Pasadena, USA, 91125} 

\correspondingauthor{Florian Debras}
\email{florian.debras@irap.omp.eu}

\begin{abstract}
Interior models of giant planets traditionally assume that at a given radius (i.e. pressure) the density should be larger than or equal to the one corresponding to a homogeneous, adiabatic stratification throughout the planet (referred to as the 'outer adiabat').  The observations of Jupiter's gravity field by Juno combined with the constraints on its atmospheric composition appear to be incompatible with such a profile.  In this letter, we show that the above assumption stems from an incorrect understanding of the Schwarzschild-Ledoux criterion, which is only valid on a local scale.  In order to fulfil the buoyancy stability condition, the density gradient with pressure in a non-adiabatic region must indeed rise more steeply than the {\it local} adiabatic density gradient.  However, the density gradient can be smaller than the one corresponding to the outer adiabat at the same pressure because of the higher temperature in an inhomogeneously stratified medium.  Deep enough, the density can therefore be lower than the one corresponding to the outer adiabat.  We show that this is permitted only if the slope of the local adiabat becomes shallower than the slope of the outer adiabat at the same pressure, as found in recent Jupiter models due to the increase of both specific entropy and adiabatic index with depth.  We examine the dynamical stability of this structure and show that it is stable against non-adiabatic perturbations.  The possibility of such unconventional density profile in Jupiter complicates further our understanding of the internal structure and evolution of (extrasolar) giant planets. 
\vspace{1cm} 

\hspace{6cm}Accepted in ApJL

\end{abstract}


\section{Introduction}
\label{sec:intro}
The combination of recent observational and theoretical significant advances, namely Juno's determinations of Jupiter's
high order gravity moments \citep{bolton2017,Iess2018} and numerical calculations of relevant dense matter equations of state (e.g., \citet{MH13,Chabrier2019,Mazevet2019, Chabrier2021}) 
have drastically impacted our understanding of Jupiter and thus of giant planets in general \citep{Wahl17,Debras2019}. 

Recent models with an adiabatic,
homogeneous envelope predict a solar or even subsolar heavy element
abundance in the atmosphere, a striking likely disagreement with
observations. In reality the atmosphere is supersolar, at least based on
methane \citep{Wong2004}, and could plausibly be 2 or even 3 times solar if water (the
most important heavy element contributor) is also supersolar by a
similar amount. If the right equation of state is in use, then it is
tempting to attribute the low density in the models to either
superadiabaticity combined with a stable compositional gradient or
subadiabaticity combined with a unstable compositional gradient. \citet{Debras2019} proposed the former case of superadiabaticity to
reconcile the relatively
modest values of the high-order gravitational
moments observed by Juno and the high, supersolar heavy element fraction revealed by Galileo.
This (counterintuitive) solution, however,
 raises an important issue. Indeed, according to the Schwarzschild-Ledoux criterion,
for a region to be stable against overturning convection, any departure from adiabaticity should lead to a steeper density gradient, thus implying a larger density at depth than the one of the outer adiabat.\footnote{We define the "outer" or "external" isentrope/adiabat as the density-pressure and temperature-pressure profiles Jupiter would have if the specific entropy was constant throughout the planet, at the value inferred from the Galileo and Juno observations, i.e. 1 bar  at 166 K \citep{Wong2004,Li2017}. We tested that choosing 170 or even 175K does not change the overall picture \citep{Leconte2017,Guillot2020-1,Guillot2020-2}. }
Therefore, a structure of Jupiter less
dense with depth than the external adiabat needs to be justified on physical grounds, 
both in terms of stability and of a plausible cooling history. 

In this paper, we derive simple physical arguments to assess the validity of such a density profile.
First, we recall in \S\ref{sec:stratif} the required conditions for a non adiabatic region
to persist in a planet's interior. In \S\ref{struc}, we show that the the local adiabatic density gradient in the non adiabatic region must become flatter than the density gradient of the outer adiabat at the same pressure. We show that this condition is fulfilled in recent Jupiter models because of H$_2$ pressure dissociation and atomic He enrichment at the expense of molecular H$_2$, which both yield a decrease of the number of degrees of freedom and therefore an increase of the adiabatic index.
Then, in \S\ref{sec:stability} we examine the dynamical stability of such density structures.
We briefly examine the type of non adiabatic evolution for the planet in \S\ref{sec:evolution}.
Section \ref{sec:conclusion} is devoted to the conclusion. 
In spite of their relative simplicity, that need further numerical simulations of 
much greater complexity to be fully validated, these calculations provide physically sound arguments
to justify the possibility of non adiabatic temperature and density stratifications in Jupiter and giant planet interiors.

\section{Conditions for a non adiabatic region}
\label{sec:stratif}

For a given composition, associated with a mean molecular weight $\mu$, the thermodynamic properties of the fluid can be calculated from the knowledge of two of the following quantities, specific entropy $S$,  temperature $T$,  pressure $P$ and  mass density $\rho$. Notably, $\rho$ can be calculated from $P$ and $S$ to derive Jupiter models with the concentric MacLaurin spheroid method \citep{Hub2013,Debras2018}.

Under Jupiter conditions, where motions are small compared to the speed of sound, an adiabat can be approximated as an isentrope
($dQ=0\Longleftrightarrow dS=0$),
at least in the absence of irreversible processes such as e.g. phase separation. 
In regions where the molecular weight is constant the temperature gradient in the planet is thus equal to the adiabatic temperature 
gradient (\citet{Hubbard1968,Saumon1992_2,Hubbard2002}), defined 
as:

 \begin{equation}
 \nabla_{T_{(\mu = cte)}} \equiv \left(\dfrac{\mathrm{d} \ln T}{\mathrm{d} \ln P}\right)_{\mu = cte} \simeq  
 \left(\dfrac{\mathrm{d} \ln T}{\mathrm{d} \ln P}\right)_{\mathrm{ad}} \equiv \nabla_{T_{ad}}.
 \end{equation}
As in convective regions the specific entropy is constant, this is equivalent 
 to say that the density gradient must obey the condition:

 \begin{equation}
\nabla_{\rho_{(\mu = cte)}} \equiv \left(\dfrac{\mathrm{d} \ln \rho}{\mathrm{d} \ln P}\right)_{\mu = cte} \simeq  \left(\dfrac{\mathrm{d} \ln \rho}{\mathrm{d} \ln P}\right)_{\mathrm{ad}} \equiv \nabla_{\rho_\mathrm{ad}} .
\end{equation}
The need for a lower density than that of an 
adiabatic profile requires that, somewhere within the planet, $\nabla_\rho < \nabla_{\rho_\mathrm{ad}}$, which implies regions of compositional gradients.
A regime of double-diffusive instability may then develop due to the competition between (fast) thermal diffusivity and (slow) molecular diffusivity. Such regions are stable to overturning convection, thus stable with respect to the Ledoux criterion, but still unstable to small-scale convection, i.e. unstable w.r.t. the Schwarzchild criterion. Two situations can occur depending on
the destabilizing or stabilizing nature of the compositional gradient: 


\begin{enumerate}
\item in the former case, the molecular weight is decreasing with depth in some part of the planet ($\mathrm{d} \mu/ \mathrm{d} r > 0$, i.e. $\nabla_\mu= \mathrm{d} \ln \mu / \mathrm{d} \ln P\textless 0$) and can lead to a region of fingering convection; 

\item in the second, opposite case, the increasing
molecular weight with depth ($\mathrm{d} \mu/ \mathrm{d} r < 0$, i.e. $\nabla_\mu\textgreater 0$)
can trigger oscillatory or, more likely, layered convection, possibly under the form of blurred double diffusive convection,
a regime generally identified as "semi convection" (\citet{Mirouh2012,Moll2016}). 
\end{enumerate}

These two situations can be described by the two following respective conditions (e.g. \citet{Rosenblum2011}):

\begin{eqnarray}
\text{fingering convection:   } &1& < R_\rho < \frac{1}{\tau}, \label{ineq:DD_T1}\\
\text{semi-convection:   } &1& < R_\rho^{-1} < \frac{Pr+1}{Pr+\tau},
\label{ineq:DD_T2}
\end{eqnarray}
where $Pr=\nu/\kappa_T$ and $\tau=\kappa_\mu/\kappa_T$ denote respectively the Prandtl and inverse Lewis numbers, with $\nu$ the
molecular viscosity, $\kappa_\mu$ and $\kappa_T$ the molecular and thermal diffusivities, $R_\rho=\frac{\alpha_T}{\alpha_\mu}\frac{\nabla_T-\nabla_{T_{ad}}}{\nabla_\mu}$ is the density contrast parameter and

\begin{eqnarray}
\alpha_\mu &=& \left( \dfrac{ \partial \ln \rho}{\partial \ln \mu} \right)_{T,P}=
-\left( \dfrac{ \partial \ln \rho}{\partial \ln P} \right)_{T,\mu} \left( \dfrac{ \partial \ln P}{\partial \ln \mu} \right)_{T,\rho} \\
\alpha_T &=& -\left( \dfrac{ \partial \ln \rho}{\partial \ln T} \right)_{P,\mu}=
\left( \dfrac{ \partial \ln \rho}{\partial \ln P} \right)_{T,\mu} \left( \dfrac{ \partial \ln P}{\partial \ln T} \right)_{\rho,\mu}
\label{def_alpha}
\end{eqnarray}
only depend on the equation of state (EOS). For any realistic EOS in Jupiter, $\alpha_T > 0$ and $\alpha_\mu > 0$.

These conditions can be rewritten in terms of density, instead of temperature. For a general EOS
$P(\rho,T,\mu)$, it is easy to show that these conditions become:

\begin{align}
&\text{fingering convection:   }  \nonumber \\
\nabla_{\rho_\mathrm{ad}} < \nabla_\rho& < \nabla_{\rho_\mathrm{ad}}+\left(\frac{1}{\tau}-1\right)\frac{\chi_\mu}{\chi_\rho}\nabla_\mu,
\label{ineq:DD_rho2} \\
&\text{semi-convection:   } \nonumber \\
\nabla_{\rho_\mathrm{ad}}<\nabla_\rho& < \nabla_{\rho_\mathrm{ad}}-\left(\frac{1-\tau}{1+Pr}\right)\frac{\chi_\mu}{\chi_\rho}\nabla_\mu,
\label{ineq:DD_rho1}
\end{align}
where 
\begin{align}
&\chi_\mu = \left( \dfrac{ \partial \ln P}{\partial \ln \mu} \right)_{T,\rho},
&\chi_\rho = \left( \dfrac{ \partial \ln P}{\partial \ln \rho} \right)_{\mu,T}.
\end{align}
Since under the conditions of interest $\tau\approx 10^{-2}\ll1$, $Pr\approx 10^{-2}$-$10^{-1}$ (see e.g. \citet{Chabrier2007}), $\chi_\mu<0$, $\chi_\rho>0$ and $\nabla_\mu$ respectively $<0$ and $>0$ in the fingering and semi-convection case, these inequalities can be fullfilled. 

The left hand side of the two inequalities expresses the same property:
in a system stable to overturning convection, the density of a parcel of fluid raised adiabatically to a lower pressure must be larger than that of its new surroundings.
The right hand side conditions correspond to the stratification limits below which
the regions are stable
w.r.t the Ledoux criterion but unstable to small-scale double-diffusive instability.


A region of
fingering convection in the outermost layers, i.e. $\nabla_\mu< 0$, , may occur either because of (i) accretion of hot entropy gas during the
runaway gas accretion phase,  hampering or even completely inhibiting convection in the outermost part of the planet \citep{Berardo2017,BerardoC2017} and/or (ii) the coaccretion of rock and gas or accretion of planetesimals
over some fraction of the planet's surface during the planet's history (e.g. \citet{Iaro2007,Mordasini2017}). 


In this paper, we only consider a region of increasing
molecular weight with depth ($\mathrm{d} \mu/ \mathrm{d} r < 0$, i.e. $\nabla_\mu\textgreater 0$)  as there  is  {\it a need} for such a region in Jupiter's interior. 
Indeed, Galileo's observations of helium atmospheric abundance \citep{vonzahn1998} reveal only 90\% of the protosolar value inferred from
solar models (e.g., \citet{Anders1989}), whereas
the global helium content of Jupiter is presumably protosolar.
This increase of helium mass fraction with depth yields an increase of molecular weight. Such a gradient could be due either to H/He immiscibility,  leading to helium sedimentation \citep{Stevenson1977,Stevenson1979,Fortney2003}, or to a region of layered or blurred double diffusive convection, possibly triggered by immiscibility (see 
the discussions in \citet{Stevenson1977,Debras2019}).
It is worth noting that
semi-convection occurs more easily in regions where $|\chi_\mu / \chi_\rho| > 1$ (see Eq.\eqref{ineq:DD_rho1}), 
hence deeper than $~1$ GPa (=10 kbar, \citet{Debras2019}).

At first sight, the conditions \eqref{ineq:DD_rho2} and \eqref{ineq:DD_rho1} seem to contradict the recent models of \citet{Debras2019} since the first obvious conclusion is that
the density gradient must be {\it steeper} than the adiabatic density gradient.
The correct analysis, however, is more subtil.
In a convective homogeneous 
medium, the adiabatic profile can be considered as a global profile and the specific entropy 
is constant throughout the entire region. In a medium prone to composition change, however, $\nabla_{\rho_\mathrm{ad}}$
is only defined on a {\it local scale}, as the specific entropy changes with depth. 
Therefore, the density gradient can be always steeper than the {\it local} adiabatic density gradient, defined by the thermodynamical properties of the medium at a given pressure, although flatter than the density gradient the outer adiabat would have at the same pressure.
This issue is explored in detail in the following sections.

In all cases, this implies an intermediate {\it inhomogeneous} region departing from an adiabatic profile, bracketed by two adiabatic ones in Jupiter's gas rich envelope. 
According to the analysis of \citet{Debras2019} using state-of-the art H/He eos (\citet{Chabrier2019, Chabrier2021}) , sufficiently deep in this region the density should be lower than the density of the external isentrope at the same pressure.

\section{Thermodynamic considerations}
\label{struc}
\begin{figure}[ht!]
\centering
  \includegraphics[width=\linewidth]{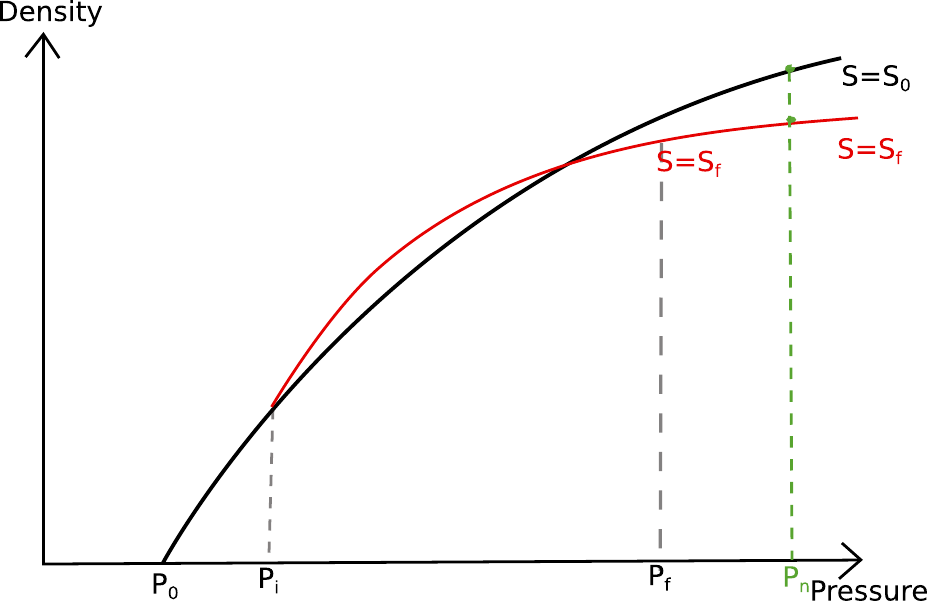}
\caption{Cartoon of the superadiabatic profile
compared with an adiabatic one. The overdense region compared to the adiabat is exagerated for illustration purposes.}
\label{fig:struct}
\end{figure}

Figure \ref{fig:struct} displays
a $\rho(P)$ profile where we suppose the existence of a
superadiabatic region between pressures $P_i$ and $P_f$, associated with a decrease in density at depth (red line), 
compared with the outer adiabatic profile (black line).  The outer and inner
specific entropies are $S=S_0$ at $P_i$ and $S=S_f$ at $P_f$, respectively.

The non adiabatic region can be characterized by two physical processes: a diffusive region or a semi-convective region. In both cases, the left hand side of Ineq.\eqref{ineq:DD_rho1} must be verified: the density gradient must be steeper than the density gradient of  the local adiabat. However, Figure \ref{fig:struct} shows that we want the density gradient to be shallower than the one of the outer adiabat. Therefore, the non-adiabatic region must verify:

\begin{equation}
     \left.\dfrac{\mathrm{d} \rho}{\mathrm{d} P}\right|_S < \dfrac{\mathrm{d} \rho}{\mathrm{d} P} < \left.\dfrac{\mathrm{d} \rho}{\mathrm{d} P}\right|_{S_0}, \label{ineq:gradient}
\end{equation}
where $\left.{\mathrm{d} \rho}/{\mathrm{d} P}\right|_S$ is the adiabatic density gradient at pressure P in the region, ${\mathrm{d} \rho}/{\mathrm{d} P}$ is the actual density gradient and $\left.{\mathrm{d} \rho}/{\mathrm{d} P}\right|_{S_0}$ is the density gradient of the outer adiabat at the same pressure P.

It is easily shown that for an adiabatic reversible transformation ($dQ=TdS=0$), the density gradient must verify: 

\begin{equation}
    \left.\dfrac{\mathrm{d} \rho}{\mathrm{d} P}\right|_S = \gamma^{-1} \left(\dfrac{\partial \rho}{\partial P} \right)_{T},
\end{equation}
where $\gamma=C_P/C_V$ is the usual adiabatic index, i.e. ratio of the specific heats at constant pressure and volume, respectively. For an ideal gas, this relation yields the Laplace law. For a general EOS, $\gamma$ depends on the thermodynamic quantities. As the temperature is a function of pressure and entropy in our models, Ineq.\eqref{ineq:gradient} then implies a condition on the {\it local} adiabatic density gradient $\left.{\mathrm{d} \rho}/{\mathrm{d} P}\right|_S$, namely: 
\begin{equation}
    \gamma^{-1}(P,S) \left(\dfrac{\partial \rho}{\partial P} \right)_T (P,S) < \gamma^{-1} (P,S_0) \left(\dfrac{\partial \rho}{\partial P} \right)_{T_0} (P,S_0). \label{ineq:slope}
\end{equation}

 If condition (\ref{ineq:slope}) is satisfied in the non adiabatic regions of Jupiter, the density gradient can then become flatter than the one of the outer adiabat. Eventually, the density of the non-adiabatic structure can thus become lower than the density of the outer adiabat at the same pressure. 
 Note that, because of the superadiabiticity in the semi-convection zone, each layer lies on a warmer isentrope than the outer one, $T(P,S)>T_0(P,S_0)$. This contributes to decreasing $\left(\partial \rho/{\partial P} \right)_T$.

\begin{figure}[ht]
\centering
  \includegraphics[width=\linewidth]{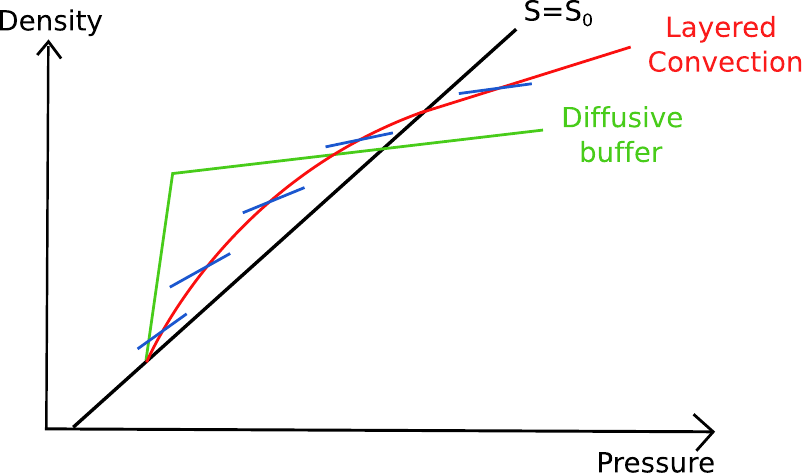}
\caption{Cartoon of P-$\rho$ profiles for: black: isentropic profile, red: semi-convection region, 
green: diffusive buffer. The blue lines are the local isentropes in the 
semi-convective profile. 
Comparing black with red or green lines, the diffusive or layered convective profiles   exhibit {\it initially} a  steeper density gradient than the isentropic profile, but  become eventually less dense at depth because of the decreasing steepness of the density gradient. 
}
\label{fig:LC_2}
\end{figure}

Such a profile is illustrated in
Figure \ref{fig:LC_2} for two cases: a single, small diffusive buffer or an extended layered convection region. 
At the bottom of the diffusive buffer, the adiabatic gradient is much flatter than the gradient of the outer adiabat because of the sharp increase in temperature. The semi-convective region, on the other hand, is characterized by a slow flattening of the local adiabatic gradients with depth. In both cases, the density eventually becomes lower than the density of the outer adiabat at depth.
\footnote{Strictly speaking, the pressure at given radius will be slightly different between the fully adiabatic and superadiabatic profiles. 
However, the difference in the two P(r) profiles,  which corresponds to the change of gravity between the two models, is very slow to develop and fairly negligible (see Fig. 9 of \citet{Debras2019}).}

\begin{figure*}[ht!]
\centering
  \includegraphics[width=0.43\linewidth]{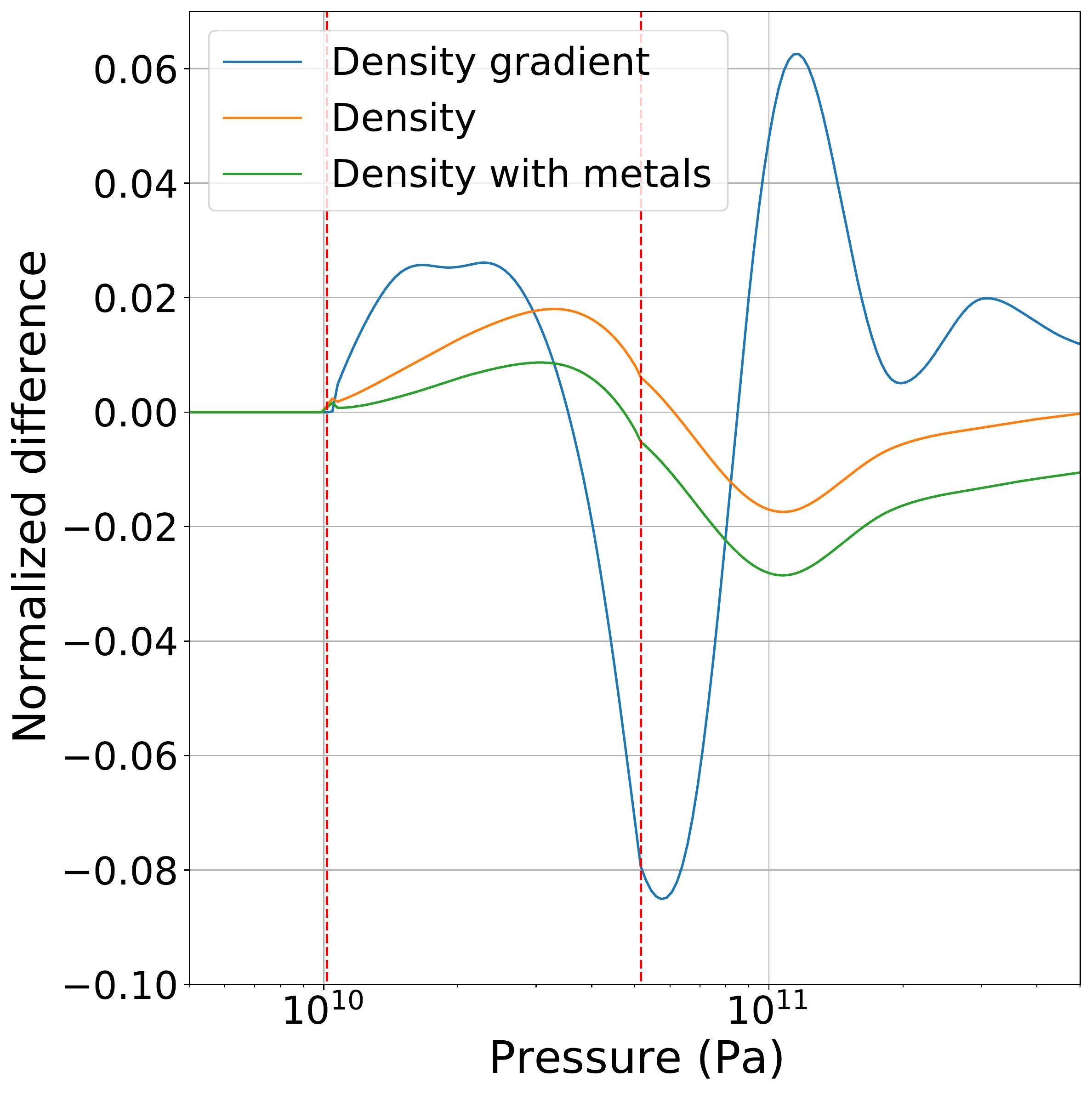}
\hspace{1cm}  
    \includegraphics[width=0.43\linewidth]{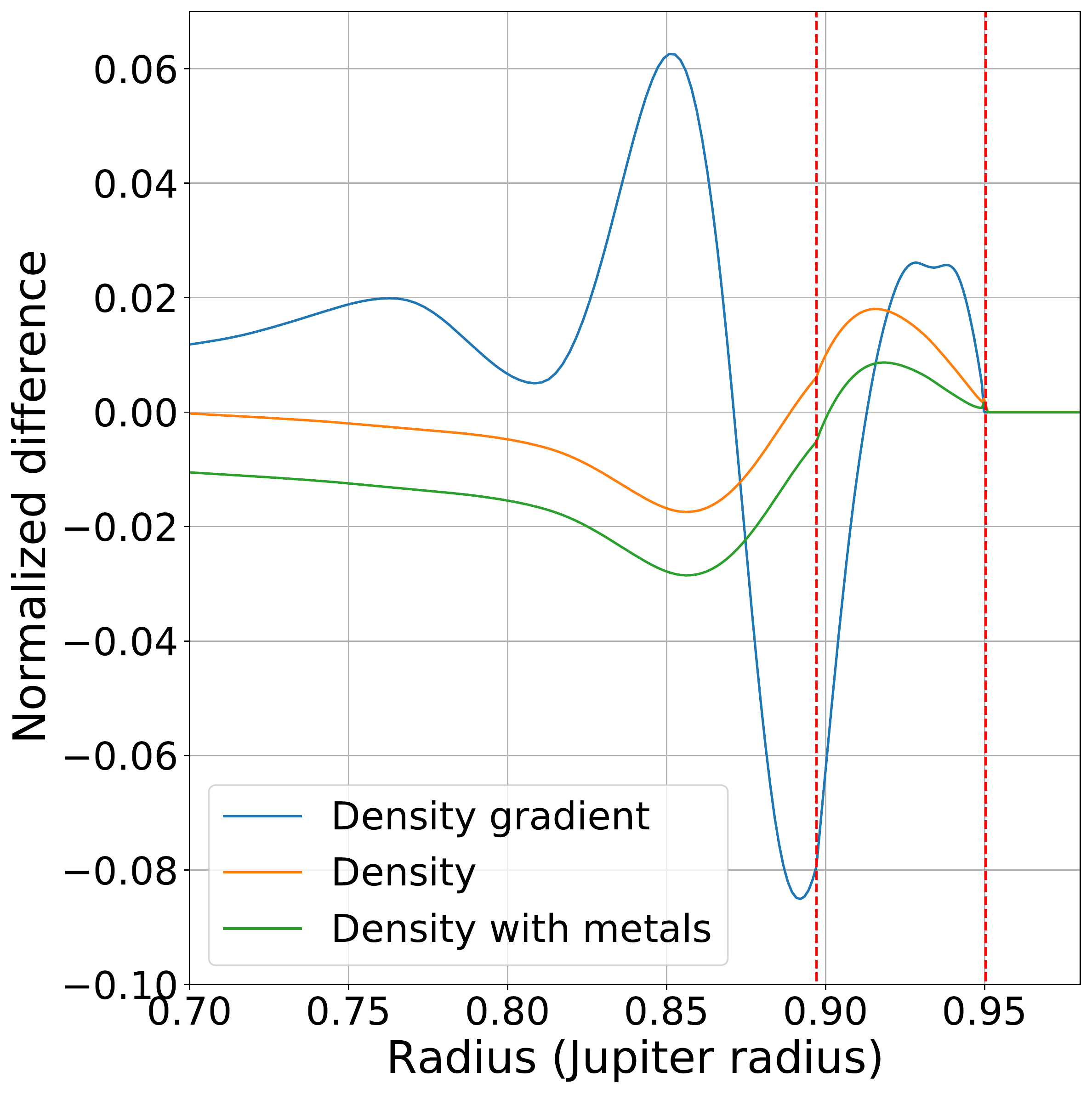}
\caption{Blue curves: difference between the local adiabatic density gradient from a model of \citet{Debras2019} and the density gradient of the outer
adiabat at the same pressure (left) or radius (right), divided by the density gradient of the outer adiabat. Orange curves: difference between density of the model and the density of the outer adiabat, divided by the density of the outer adiabat. Green curves: same as orange curve but including the decrease of metal content with depth as proposed in \citet{Debras2019}. Red vertical dashed lines: limits of the semi-convective zone, separating the two convective envelopes. 
Note: 1 Mbar = $10^{11}$ Pa.}
\label{fig:gradients}
\end{figure*}
In Figure \ref{fig:gradients}, we show the difference between the local adiabatic density gradient and the density gradient of the outer adiabat in a typical model of \citet{Debras2019}, as well as the difference between the true density and the one of the outer adiabat, as a function of pressure and radius, respectively. We see that there is a direct correlation between the two curves: when Ineq.\eqref{ineq:slope} is verified (blue curve negative), the density decreases compared to the density of the outer adiabat (orange/green curve decreasing), and inversely, confirming the validity of the model. 
{The fact that Ineq.\eqref{ineq:slope} is fullfilled in the semi-convection region of Jupiter ($P\simeq 10-100$ GPa) stems from several factors. First, as mentioned above, the superadiabaticity decreases the local density gradient, $\left({\partial \rho}/{\partial P} \right)_T$. Second, the adiabatic index $\gamma$ increases with depth because of (i) H$_2$ pressure dissociation, (ii) atomic He enrichment at the expense of molecular H$_2$, which both yield a decrease of the number of degrees of freedom.}

We see in Figure  \ref{fig:gradients} that the density is larger than the density of the outer adiabat between ~0.1 to ~0.6 Mbar, i.e. about 4000 km, which encompasses the whole semi-convective region. This result will be used in the next section. 

\begin{figure*}[ht!]
\gridline{\fig{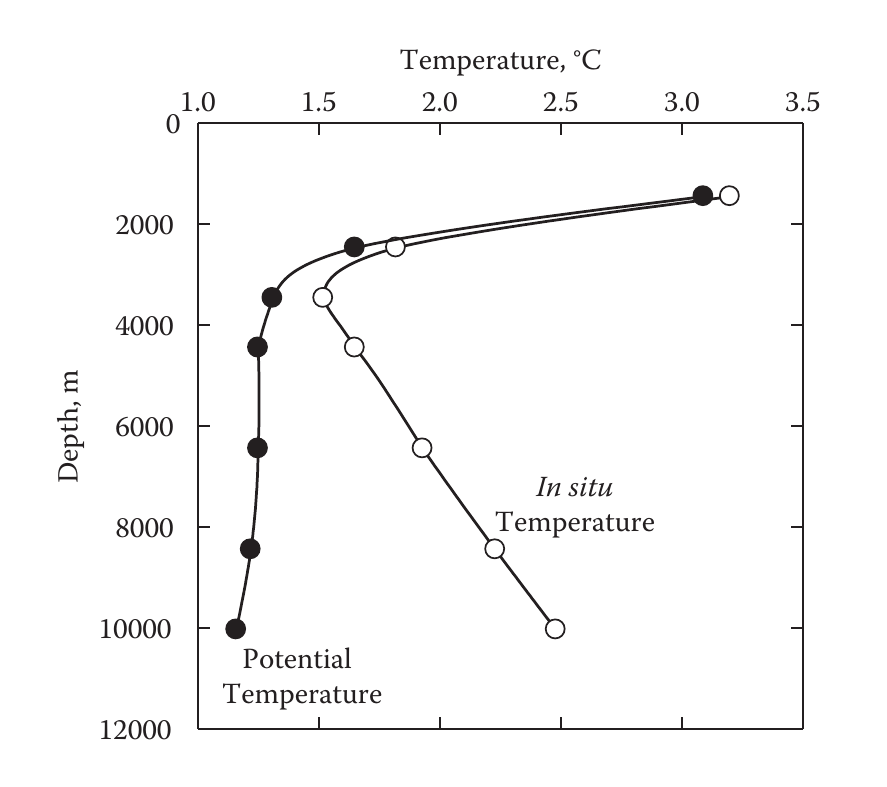}{0.5\textwidth}{}
\fig{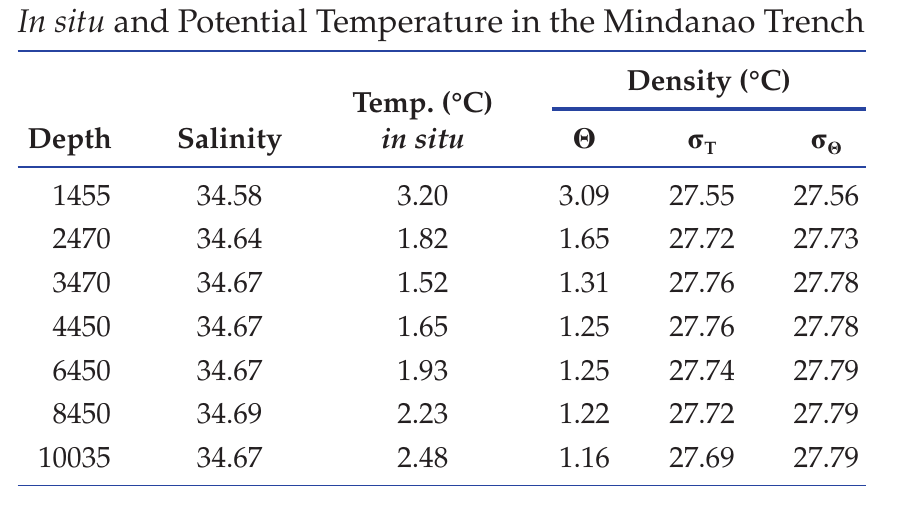}{0.5\textwidth}{}}
\caption{Salinity, {\it in situ} temperature, potential temperature ($\theta$) , {\it in situ} density ($\sigma_T$)
and potential density ($\sigma_\theta$) as a function of depth in the Mindanao trench. The 
potential temperature and density are defined here at the top of the ocean. The potential density 
increases while the actual density decreases with depth, showing the stability of a superadiabatic, less dense than an adiabat profile. 
Copyright 2013 From Chemical Oceanography, Fourth Edition by Frank Millero. Reproduced by permission of Taylor and Francis Group, LLC, a division of Informa plc.}
\label{fig:mindanao}
\end{figure*}

It is worth mentioning the existence of
this kind of density structure on Earth, namely in the Mindanao trench, 
whose thermodynamic profile is displayed in Figure \ref{fig:mindanao} (see \citet{Millero2011}). In the 
Mindanao trench, salinity and temperature are increasing with depth, salted water yielding a downward increasing molecular weight, whereas the density is decreasing 
with depth and the density at the bottom of the trench is lower than the one of an adiabat. This profile is nonetheless stable, because of the increase of potential 
density\footnote{In oceanography, the potential density and temperature are defined at the surface of 
the ocean. An increase of potential density with depth is equivalent here to a  
density gradient flatter than an adiabatic one, which can
lead to a decrease of density 
with depth, as seen in Fig.\ref{fig:mindanao}.} with depth. 
Care must be taken in the comparison as water is almost
incompressible, but this example of a steady state situation shows that a superadiabatic temperature profile associated
with a lower density than the outer adiabatic profile in a region
of downward increasing molecular weight can be stable.

\section{Conditions for dynamical stability}
\label{sec:stability}

\subsection{Buoyancy arguments}
\label{buoyancy}

In this section, we examine the stability of the pressure-density profiles mentioned in the previous section, which have been shown to be
statically stable, against dynamical perturbations.
We first examine the condition of buoyancy stability. 

Let us assume that 
a parcel of fluid from the adiabatic envelope above the semi-convective region can move downward adiabatically over great distances. 
As examined in \S\ref{sec:stratif}, static stability implies that the parcel must first travel throughout regions which exhibit steeper density gradients than the adiabatic density gradient of the parcel along its path.
A simple buoyancy argument gives the deceleration rate of the particle due to the {\it locally},  overdense surrounding medium: 

\begin{equation}
\dfrac{\mathrm{d} v}{\mathrm{d} t}  \sim \left(\dfrac{\rho_\mathrm{ad}-\rho}{\rho_\mathrm{ad}} \right) g,
\end{equation}
where $v$ is the parcel velocity, $g$ the gravitational acceleration, 
$\rho_\mathrm{ad}$ the density of the parcel displaced adiabatically and $\rho$ the density 
of the medium. For simplification, let us assume that $(\rho_\mathrm{ad}-\rho)/\rho_\mathrm{ad}$
is constant with depth, and ${\mathrm{d} v}/{\mathrm{d} t} \approx v/t_\mathrm{dec}$, with $t_\mathrm{dec}$
a characteristic deceleration timescale for the parcel. The condition for the particle to travel through a distance $l$
is roughly: 

\begin{equation}
\dfrac{l}{v} \sim t_\mathrm{dec} \Rightarrow \ \ \ 
l\sim \dfrac{v^2 \rho_\mathrm{ad}}{\left(\rho-\rho_\mathrm{ad}\right) g}.
\end{equation}
From the mixing length theory (MLT, \citet{Hansen1994,Kippenhahn1990}), one gets for the convective velocity in the $0.1-2$ Mbar region $v \sim 0.1\, \mathrm{m\, s^{-1}}$, 
and $g \sim 10\, \mathrm{m\, s^{-2}}$ for Jupiter,
while the typical departure from an adiabatic profile over a few kms in the models of \citet{Debras2019} is

\begin{equation}
\dfrac{\Delta \rho }{ \rho_\mathrm{ad}}=\left| \dfrac{\rho_\mathrm{ad}-\rho}{\rho_\mathrm{ad}} \right|_{\mathrm{50\,km}} \sim 10^{-4}.
\label{drho}
\end{equation}
An order of magnitude estimate for the distance $l$ is then:

\begin{equation}
l \sim 10\, \mathrm{m}\left(\dfrac{v}{0.1 \mathrm{m\,s^{-1}}} \right)^2 
\left( \dfrac{\Delta \rho / \rho_\mathrm{ad}}{10^{-4}}\right)^{-1} \left(\dfrac{g}{10\, \mathrm{m\,s^{-2}}} \right)^{-1}.
\end{equation}

Hence, for the non-adiabatic region to be dynamically stable, any parcel sinking adiabatically must have a lower density than the local medium over a few tens of meters. The density must thus increase more steeply with pressure than the outer adiabatic profile only over about a hundred meters or so to prevent the destabilisation of the medium.

Examining the onset of layered convection in Jupiter, \citet{Chabrier2007} and \citet{Leconte2012} analytically estimated that the typical size of 
the convective layers in the semi convection region is $\sim 100$ m.
Therefore,
for the abovementioned typical relative increase in density, of the order 
of $10^{-4}$, one single layer with a density gradient steeper than the one of the outer adiabat is sufficient
for the buoyancy restoring force to stabilize the medium.
This is of course a very crude estimate, 
but it shows that the medium is extremely stable to adiabatic perturbations. 

As mentioned in \S\ref{struc} and shown in Fig. \ref{fig:gradients},  
typical profiles in \citet{Debras2019} exhibit a higher density than the outer adiabat over the whole 
semi-convective zone, which is 3000 km large. 
This encompasses about $30,000$ convective layers, which  largely fulfills the aforementioned buoyancy stability condition. Two conclusions can then be drawn for the regions of layered convection with a density gradient shallower than the one of the external adiabat at the same pressure:

$\bullet$ the first layers at the top of the semi convection zone, where the molecular weight gradient starts to develop, must be denser than the outer adiabat at the same pressure. As shown above, one single convective layer fulfilling this condition is sufficient to stabilize the medium against buoyancy instability.

$\bullet$ for the typical departure from the adiabatic density profile found in the models (eqn.(\ref{drho})), a parcel of fluid brought  downward adiabatically from some level to higher pressure levels does not become unstable unless its velocity is  ~2-3 order of magnitudes larger than the convective velocity, because of the buoyancy restoring force.

\subsection{Wave instability}
\label{sec:RT}

Let us now consider  an adiabatic, horizontal, divergence free perturbation of given wavelength $\lambda$ and vertical amplitude $h$. We denote  $H$ the typical height over which the medium becomes less dense than the outer adiabat. The discussion of \S \ref{buoyancy} shows that $H \gtrsim 100$ m at least, and the models of \citet{Debras2019} allow for $H \sim 3000$ km. 

If the initial amplitude of the wave $h \ll H$, the wave satisfies the Schwarzschild criterion at every point: it will be damped out on a buoyant timescale. On the other hand, if $h \gtrsim H$, the antinodes of the wave are buoyantly unstable and the wave may become globally unstable. By analogy with Rayleigh-Taylor instability, rotation and magnetic field might prevent instability to occur. However,  using \citet{Chandra1961}, it can be easily verified that under Jupiter's conditions, namely angular velocity $\Omega \sim 10^{-4}\ \mathrm{s}^{-1}$ and
magnetic field amplitude of the order 
of a few tens of Gauss, $B=\mu H \approx 30$ G at $r = 0.85R_\mathrm{J}$ (\citet{Connerney2018}), i.e. at the location of the semi-convective zone \citep{Debras2019},
rotation and magnetic tension  can only stabilize short wavelength ($\lambda\lesssim 100$ km) perturbation waves. The real situation, however, is more complex than described above for three reasons: 


i) The growth timescale of the instability depends in reality on the {\it integral}
of the density difference with pressure, $\int(\rho-\rho_\mathrm{ad})v_zdP$, where $\rho_\mathrm{ad}$ is the density of the outer adiabat and
$v_z$ denotes the vertical component of the perturbation
velocity (eq.(X 44) of \citet{Chandra1961}). As the path along the integral will alternate underdense and overdense regions compared with the outer adiabat, respectively destabilizing/stabilizing the medium, it is not clear what the final result will be. At the very least, the growth timescale will be increased compared to a simple Rayleigh-Taylor estimation for two fluids of constant density.

ii) The wave must remain adiabatic over the entire superadiabatic region to have a chance to destabilize it. This
corresponds to $\sim 3000$ km in the models of \citet{Debras2019}. The perturbation must remain adiabatic over this scale since instability can occur only if the growth timescale is shorter than heat transfer timescales. Whereas atomic thermal diffusivity is too small to yield strong departure from adiabaticity ($\kappa_T\approx 10^{-1}$ cm$^2$ s$^{-1}$ for metallic hydrogen (\citet{Stevenson1977_2}),
this is less clear in the case of semi convection, with significantly enhanced diffusivity and the formation of strong localized updrafts and downdrafts \citep{Rosenblum2011}. This is even more true in case of a phase separation with a release of latent heat, i.e. $\Delta S \ne 0$.

iii) An obvious question is what physical mechanism
can excite such long wavelength perturbations and what is the typical timescale to excite them? If the perturbation is triggered by convection, the most likely hypothesis, the size of the overshoot plumes at the interface of the double-diffusive region must be 
comparable to a significant fraction of the size of this latter, i.e. $\sim 3000$ km. Assuming
rough equipartition between the wave energy and the excitation mechanism (at least in the linear phase), this will roughly
correspond to the wave amplitude, which by itself challenges the existence of such "perturbations". 
Given the typical convective
velocity $v\simeq 0.1\, \mathrm{m\, s^{-1}}$, it will take about 300 days to generate such a wave. As examined in \S\ref{buoyancy}, damping due to buoyancy occurs
on a much shorter timescale.
The problem can be rephrased the other way around:  
there is no obvious physical mechanism susceptible to excite from linear perturbations long wavelength {\it adiabatic} perturbations with an initial amplitude 
of thousands of kilometers at the interface 
between the external envelope and the inhomogeneous zone in Jupiter's structure models proposed by \citet{Debras2019}.

\section{Evolution}
\label{sec:evolution}


It is well known that an ongoing adiabatic contraction of Jupiter
since its formation yields an 
age in good agreement with the age of the solar system, as shown originally by \citet{Hubbard1977}.
Carrying out cooling calculations in case of a non adiabatic internal structure can not be done analytically
and requires numerical calculations. Although several attempts have been made to explore this issue,
none of them
has properly included {\it all} the various processes potentially responsible for such a departure from adiabaticity,
namely (i) phase separation, (ii) double diffusive convection and (iii) core erosion since recent Jupiter models strongly suggest a diluted core for Jupiter \citep{Wahl17,Debras2019}. 
Given the complexity of these processes, providing a {\it robust} complete  cooling history of Jupiter appears
to be rather elusive for now, in spite of some claims in the literature. Indeed, besides the uncertainty in existing H/He phase diagram calculations,
by itself a major source of uncertainty, there is a trade off between these various mechanims, which can
either increase or decrease the cooling rate of the planet, depending whether, and when, they lead to a
production or a loss of energy. A typical exemple, for instance, are the calculations of
\citet{Leconte2013}, that show that the correct age and luminosity of
Saturn can be obtained in the case of layered convection.
The reason is that a non adiabatic region decreases the output heat flux between the interior and the photosphere compared with an entirely convective planet,
yielding a decrease of the luminosity. As the energy transport is less efficient, however,
a non adiabatic planet eventually becomes more luminous than an adiabatic one,
as it cools down over longer timescales.
For Saturn, in the case of the calculations of \citet{Leconte2013}, the crossover happens after a few hundred
millions years (their figure 2). Layered convection or immiscibility happening about a few hundred
millions years ago in Jupiter's interior, leading to a non isentropic internal region, could thus very well be consistent
with the planet's current luminosity. Not mentioning, again, the impact of core erosion on the energy balance (\citet{Stevenson1985,Guillot2004}).

\begin{figure}[ht!]
\centering
  \includegraphics[width=\linewidth]{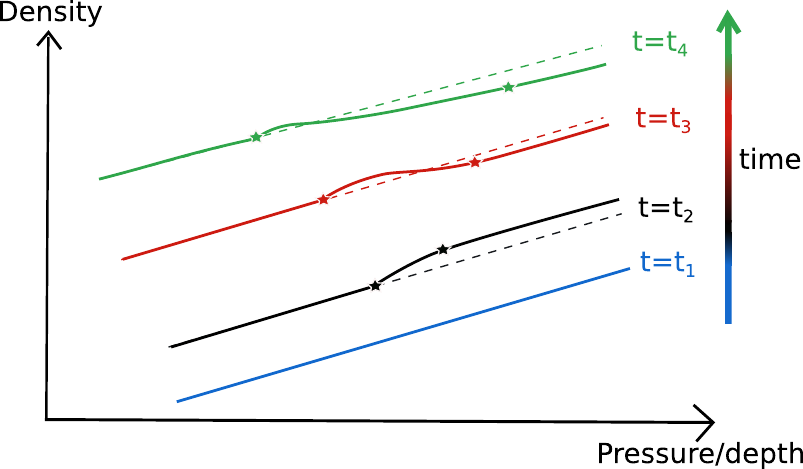}
\caption{Possible evolutionnary path for Jupiter, as outlined in \S\ref{sec:evolution}.
Straight, dashed lines correspond to fully isentropic profiles, curved solid lines illustrate
the non adiabatic pressure-density structure. The stars bracket the 
non adiabatic zone in the Mbar region. At $t=t_1$ (blue), Jupiter 
is fully isentropic, dashed and plain lines are identical. At $t=t_2$ (black), the planet has 
cooled down, immiscibility of hydrogen
and helium or semi convection has begun, Rayleigh-Taylor instability requires that 
the inner density profile must be denser than the adiabatic one. At $t=t_3$ (red), the non adiabatic 
region has expanded, and the superadiabatic temperature profile has steepened, 
leading to a lower density at higher depth than the external isentrope. At $t=t_4$ (green), 
the overdense region at the top of the semi-convection region has shrunk and 
almost the whole internal envelope exhibit a density smaller than the outer adiabat would have. 
The dilution of the core, however, might 
lead to a higher density than the adiabatic planet in the innermost part of the planet (see \citet{Debras2019}).
At this stage, the luminosity of the adiabatic and non adiabatic planets are comparable. }
\label{fig:evolution}
\end{figure}

A plausible, {\it qualitative} evolutionnary path for the planet, consistent with both its present luminosity and  
stability conditions, could thus be as follows, as shown schematically in Figure \ref{fig:evolution}:

\begin{enumerate}
	\item Formation of Jupiter as a compact core and a convective envelope. 
	Whether the original envelope is well homogeneized or not is not really consequential, as
	inhomogeneous regions could occur during the formation process or later on during the evolution (see \S2).
	\item After a certain time, the temperature in the Mbar region
	is sufficiently low for helium immiscibility to occur, possibly with other elements as well. Helium dropplets begin to drown, depleting 
	the outer envelope in helium and possibly enriching it modestly in heavy elements \citep{Stevenson1977,Debras2019}. As examined
	in \S3, initially, the inner density profile 
	must be denser than the adiabatic profile to ensure stability. The demixing process very likely triggers
	a double diffusive instability, either as a steady state or constantly dynamically generated by gravity waves at the interface with the convective envelope. Double diffusive convection, however, can very well occur before (i.e. in the absence of) H/He demixion, notably if
	external impacts hampered the convective efficiency at some epoch during Jupiter's lifetime. Once an external non-adiabatically stratified region has developed, splitting the outer envelope into two separated convective
zones, accretion of small solid bodies by Jupiter during its subsequent evolution could lead to the observed supersolar abundance
of heavy elements, whatever the initial impact of such a semi-convective zone upon the heavy elements. Accreting between 0.3 and 1.5 Earth mass of heavy elements, depending on various assumptions, over Jupiter's cooling history
would fullfill this constraint, which seems to be a plausible hypothesis (see also \citet{Podolak2020})
	\item Either process yields the presence of an extended inhomogeneous (non adiabatic) region around the Mbar level or so. The outer envelope is depleted in helium and enriched in
	heavy elements, and cools efficiently. In contrast, the inner envelope, slightly enriched
	in helium, cannot efficiently evacuate its heat content and thus develops
	a superadiabatic temperature profile, as discussed in \S \ref{sec:stability}. The luminosity 
	of Jupiter is then lower than for a purely adiabatic contraction. As examined in this paper, the density
	in the superadiabatic region can  be smaller than the density the outer
	adiabatic profile would have at the same radius, a condition
	that seems to be required to fullfill the gravity field constraints \citep{Debras2019}. 
	\item At later stages, the outer envelope has cooled 
	further down and then the heat flux of the inner envelope is released more efficiently: the 
	planet then cools more quickly than a fully adiabatic planet (see Figure 2 of \citet{Leconte2013}).
	Adding up somehow the work required to dilute the core at the expense of convection, the non adiabatic planet can eventually have a luminosity today 
	consistent with Jupiter's present luminosity.
\end{enumerate}

\section{Conclusion}
\label{sec:conclusion}
In this work, we have explored the consequences of the onset of non adiabatic temperature
 and density stratifications in Jupiter and, more generally, in gaseous giant planet interiors. 
We have explored the possibility to decrease the steepness of the density gradient with pressure compared with the one of the adiabat of uniform composition. A non adiabatic region is necessarily characterized by a change in the mean molecular weight $\mu$ with depth, and we have focused on the case of semi-convection, where $\mu$ increases with depth.

On a local scale, we have shown that the density gradient must always be steeper than the {\it local} adiabatic density gradient. However, if the slope of the adiabatic density gradient
decreases with depth in the non adiabatic region, deep in the planet this gradient can become flatter than the one the outer adiabat would have at the same pressure. The density at depth can then very well be lower than the density of the outer adiabat at the same pressure. Such a structure requires a decrease of $\gamma^{-1} \left(\partial \rho/\partial P \right)_T$  compared with the outer adiabat (see Eq.(\ref{ineq:slope})), a condition fulfilled in typical models of \citet{Debras2019}. This decrease arises essentially from the superadiabatic temperature stratification and, potentially, from
hydrogen pressure dissociation and/or hydrogen-helium phase separation.


We have shown that the medium is statically stable, and dynamically stable against non-adiabatic or low amplitude adiabatic perturbations. Only adiabatic perturbations of large initial amplitude ($\gtrsim 1000$ km) could destabilize the medium, but we did not  find a plausible origin and a physical justification for the existence of such perturbations.

Although hydrostatic calculations do not enable us to
explore in detail such dynamical considerations,
the present analysis provides constraints at the km-scale on Jupiter's inner stucture.
Interestingly enough, the existence of a similar profile on Earth in the Mindanao trench, where the molecular weight 
and temperature increase with depth whereas the density decreases, confirms the long term stability of such a peculiar density profile.

Finally, we have shown that  the planet evolution does not provide
strong enough constraints on the inner structure profile. Indeed, given the complexity 
of the various possible physical processes in the interior of gaseous planets (helium rain, semi-convection, core dilution, etc...), and our ignorance
of their proper description, reliable evolutionary calculations for the planet remain out of reach for now.

All in all, this paper demonstrates the possibility of stable 
superadiabatic regions in Jupiter and giant planets, where the density is smaller than the 
density the outer adiabat would have at the same radius.
Although numerous questions remain open to fully validate these models, 
high quality observations coupled with state-of-the-art equations of state for dense matter have drastically changed our understanding of giant planet structure and evolution, with major
consequences for extrasolar planets in general. Notably, the present calculations confirm that: 
\begin{enumerate}
\item there is {\it no 
correlation between the observed external
abundance of heavy elements and the planet's bulk composition} as convection can be hampered early in the planet's history,
\item the cooling
of gaseous planets can be much more complex than the standard paradigm of homogeneous, adiabatic contraction.
\end{enumerate}

\acknowledgements

The authors are thankfull to Christoph Mordasini, Etienne Jaupart for helpful conversations. We are also grateful to the referee whom detailed report greatly helped us  improving  the manuscript. This work was
supported by the Programme National de Planetologie (PNP) of CNRS-INSU co-funded by CNES. FD thanks the
European Research Council (ERC) for funding under the
H2020 research \& innovation programme (grant agreement
\#740651 NewWorlds).

\bibliography{biblio_Jup}

\end{document}